\documentclass[11pt]{article}
\usepackage[hmargin=1in, vmargin=1in]{geometry} 
\usepackage{amsmath}
\usepackage{amssymb}
\usepackage{amsthm}
\usepackage{bbm}
\usepackage{graphicx}
\usepackage{natbib}
\usepackage[multiple]{footmisc}
\begin{document}
\newcommand{\indicator}[1]{\mathbbm{1}\left[ {#1} \right] }
\newcommand*\wrapletters[1]{\wr@pletters#1\@nil}
\def\wr@pletters#1#2\@nil{#1\allowbreak\if&#2&\else\wr@pletters#2\@nil\fi}


\title{Principal component gene set enrichment (PCGSE)}
\author{H. Robert Frost
\footnote{Institute for Quantitative Biomedical Sciences, Geisel School of Medicine, Lebanon, NH 03756} 
\footnote{Section of Biostatistics and Epidemiology, Department of Community and Family Medicine, Geisel School of Medicine, Lebanon, NH 03756} 
\footnote{Department of Genetics, Dartmouth College, Hanover, NH 03755}, Zhigang Li$^{*\dagger}$ and Jason H. Moore\,$^{*\dagger\ddag}$}


\maketitle

\begin{abstract}

\noindent \textbf{Motivation:} Although principal component analysis (PCA) is widely used for the dimensional reduction of biomedical data, interpretation of PCA results remains daunting. Most existing methods for interpreting principal components (PCs) attempt to explain each PC in terms of a small number of variables by generating approximate PCs with few non-zero loadings. Although these methods are useful when just a few variables dominate the population PCs, they are often inadequate for characterizing the biological signal represented by the PCs of high-dimensional genomic data. For genomic data, reproducible and biologically meaningful PC interpretation requires methods based on the combined signal of functionally related sets of genes.  While gene set testing methods have been widely used in supervised settings to quantify the association of groups of genes with clinical outcomes, these methods have seen only limited application for testing the enrichment of gene sets relative to sample PCs.

\noindent \textbf{Results:} We describe a novel approach, principal component gene set enrichment (PCGSE), for computing the statistical enrichment or depletion of gene sets relative to PCs computed from genomic data. 
The PCGSE method performs a two-stage competitive gene set test using the correlation between each gene and each PC as the gene-level test statistic with flexible choice of both the gene set test statistic and the method used to compute the null distribution of the gene set statistic.
Using simulated data with simulated gene sets and real gene expression data with curated gene sets, we demonstrate that biologically meaningful and computationally efficient results can be obtained from a simple parametric version of the PCGSE method that performs a correlation-adjusted two-sample t-test between the gene-level test statistics for gene set members and genes not in the set.

\noindent \textbf{Availability:} http://cran.r-project.org/web/packages/PCGSE/index.html

\noindent \textbf{Contact:} rob.frost@dartmouth.edu or jason.h.moore@dartmouth.edu

\end{abstract}

\section{Introduction}

Developed independently by Karl Pearson \citep{Pearson:1901fk} and Harold Hotelling \citep{Hotelling:1933uq}, PCA is a well established statistical technique that performs a linear transformation of multivariate data into a new set of variables, the principal components (PCs), that are linear combinations of the original variables, are uncorrelated and have sequentially maximum variance \citep{jolliffe2002principal}. 
The solution to PCA is given by the spectral decomposition of the covariance matrix with the variance of the PCs specified by the eigenvalues, arranged in decreasing order, and the PC directions specified by the associated eigenvectors. 

In the biomedical domain, PCA has been extensively employed for the analysis of genomic data including measures of DNA variation, DNA methylation, RNA expression and protein abundance \citep{Ma:2011kl}. Common features of these datasets, and the motivation for spectral decomposition methods, are the high dimensionality of the feature space (i.e., from thousands to over one million), comparatively low sample size (i.e., $p \gg n$) and significant collinearity between the features. The most common uses of PCA with genomic data involve dimensionality reduction for visualization \citep{alter_singular_2000, Hibbs:2005fk}) or clustering of the observations \citep{yeung_principal_2001}, with population genetics an important use case \citep{Patterson:2006uq}. PCA has also been used as the basis for feature selection \citep{Lu:2011ys}, gene clustering \citep{Hastie:2000fk} and bi-clustering \citep{kluger_spectral_2003}. More recent applications include dimensionality reduction prior to gene set testing \citep{Tomfohr:2005bv, Kong:2006jb, ma_identification_2009, bruckskotten_pca2go:_2010, Chen:2011dq} and high-dimensional regression  \citep{Hastie:2009qf}.

Although PCA is a popular and effective tool for reducing the dimensionality of genomic data, application of the method remains limited by the challenge of biological interpretation \citep{Zou:2006bs, Ma:2011kl}. Because PCs are linear combinations of all original variables, which can number from the thousands to the millions for genomic data sets, they typically lack any clear biological meaning. While PCA may improve the performance of many statistical methods, e.g., better predictive accuracy in a regression context, the underlying model is often a black box.

Approaches for generating more interpretable PCs have evolved from component thresholding \citep{jolliffe2002principal}, simple components (i.e., PC loading vectors constrained to values from $\{-1,0,1\}$) \citep{vines2000} and rotation techniques (e.g., varimax) \citep{jolliffeRotPCA1995} to sparse PCA methods, which compute approximate PCs using cardinality \citep{moghaddam2006spectral, daspremont_direct_2007, Sriperumbudur:2011oq} or LASSO-based \citep{jolliffeSPCA2003, Zou:2006bs, shen2008spca, Witten:2009tg} constraints on the component loadings. By generating approximate PCs with few non-zero loadings, all of these techniques improve interpretability by associating only a small number of variables with each PC. While such sparse PCA methods can be very effective when the true population PCs are associated with only a few variables, they will fail to accurately estimate the spectral structure of the data when the population PCs are defined by the coordinated action of large groups of variables with small marginal effects.  For genomic data, the pathway-based patterns that dominate the robust structure of genetic associations with clinical phenotypes \citep{Allison:2006nx}, and are the motivation for traditional gene set testing methods \citep{huang_bioinformatics_2009, Khatri:2012fk}, can be expected to also characterize the PCs of those data sets. The PCs of genomic data are therefore more likely to be quantitatively described, in a repeatable fashion, by collections of functionally related genes, e.g., gene sets from the Gene Ontology (GO) \citep{ashburner_gene_2000} or pathways from the Kyoto Encyclopedia of Genes and Genomes (KEGG) \citep{kanehisa_kegg:_2000}, than by individual genes.

To support interpretation of PCs in terms of \textit{a priori} variable groups, rather than just individual variables, sparse PCA methods have recently been extended to include structured sparse penalties \citep{jenatton_structured_2010, grbovic2012sparse}, such as the group lasso  \citep{yuan2006grouplasso, 2010arXiv1001.0736F} and overlapping group lasso \citep{2010arXiv1009.0306L, 2011arXiv1110.0413O}. Although structured sparse PCA techniques generate sparse PC loading vectors that reflect group structure, these methods cannot be easily used to compute the statistical association between variable groups and each PC in such a way that the variable groups can be ranked according to deviation from a specific null hypothesis, as is done in traditional gene set testing.
Matrix correlation methods, such as Yanai's GCD \citep{Yanai:1980pt, jolliffe2002principal, Ramsay:1984fk}, have also been used to quantify the association between groups of variables and one or more PCs. 
However, because such matrix correlation methods compute the association of each variable group independent of the variables that do not belong to the group, they can only be used for self-contained gene set tests \citep{Goeman:2007pr} ($Q_2$ in the terminology of Tian et al. \citep{Tian:2005zr}) in a manner similar to Goeman and Buhmann's \textit{globaltest} \citep{Goeman:2004vn} and not for competitive gene set testing ($Q_1$ in the terminology of Tian et al.).

To date, competitive gene set testing relative to PCs has been limited to methods, such as Fisher's Exact Test, that are based on a $2 \times 2$ contingency table representing the association between gene set membership and a discretization of the ranked list of PC loading values \citep{roden_mining_2006}. Such contingency table tests have two key flaws: they rely on an arbitrary threshold of the gene-level test statistic, which reduces statistical power and, more importantly, they are based on the incorrect assumption of independence among the gene-level test statistics, causing them to generate high type I error rates \citep{Goeman:2007pr, Barry2008, Wu:2012fk}. These same flaws apply equally in the context of gene set testing relative to PCs. Because of the anti-conservative nature of contingency table-based tests, and other approaches that assume independence among gene-level test statistics under the null, the use of these methods for standard gene set testing has been strongly discouraged in favor of techniques that preserve inter-gene correlation, usually via permutation of the sample labels \citep{Goeman:2007pr}. Competitive gene set testing methods that correctly account for correlation among gene-level test statistics, either through sample permutation, parametric approximation of the sample permutation distribution or correlation adjustment of parametric test statistics, include SAFE \citep{Barry:2005kx, Barry2008, Zhou:2013ys}, GSEA  \citep{subramanian_gene_2005}, GSA  \citep{Efron:2007uq} and CAMERA \citep{Wu:2012fk}.

Although biologically meaningful and repeatable interpretation of the PCs of genomic data requires approaches based on functional gene sets, researchers do not currently have access to methods that competitively test the association between gene sets and PCs with correct handling of inter-gene correlation to control type I errors. To address this gap, we have developed principal component gene set enrichment (PCGSE), an approach for interpreting the PCs of genomic data via two-stage competitive gene set testing in which the correlation between each gene and each PC is used as a gene-level statistic with flexible choice of both the gene set test statistic and the method used to compute the null distribution of the gene set statistic. Although described in the context of functional gene sets and genomic data, the PCGSE method can be used to compute the statistical association between any collection of variable groups and the PCs of an empirical dataset. To enable the easy application of the PCGSE method by other researchers, we implemented the \textit{PCGSE} R package, which can be downloaded from the CRAN repository. Using simulated data with simulated gene sets and real gene expression data with curated gene sets, we demonstrate that biologically meaningful and computationally efficient results can be obtained from a simple parametric version of the PCGSE technique, based on the CAMERA method \citep{Wu:2012fk}, that performs a correlation-adjusted two-sample t-test between the gene-level test statistics for gene set members and genes not in the set.

\section{Methods}
\subsection{PCGSE inputs}
The PCGSE method takes the following data structures as input:

\begin{enumerate}		
\item \textit{Matrix of genomic data:} 
$n \times p$ matrix $\mathbf{X}$ quantifying $p$ genomic variables under $n$ experimental conditions, e.g., mRNA expression levels measuring using microarray technology. This data will be modeled as a sample of $n$ independent observations from a $p$-dimensional random vector $\mathbf{x}$. Although PCGSE does not have specific distributional requirements, sources of genomic data, especially gene expression data, are typically well represented by a multivariate normal distribution $\sim \mathcal{N}(\mu_{p \times 1}, \Sigma_{p \times p})$, especially after appropriate transformations. Often, $p \gg n$. 
	\begin{equation} \label{eqn:X}
	 \mathbf{X} = \begin{bmatrix}
	  x_{1,1} & \cdots & x_{1,p} \\
	  \vdots  & \ddots & \vdots  \\
	  x_{n,1} & \cdots & x_{n,p} \\
	 \end{bmatrix} 
	\end{equation}
\noindent where $x_{i,j}$ represents the abundance of genomic variable $j$ under condition $i$. It is assumed that any desired data transformations (e.g., log transformation of mRNA expression ratios, etc.) have been performed and that missing values have been imputed or removed for a complete case analysis. 

\item \textit{Matrix of functional annotations:} 
$f \times p$ binary annotation matrix $\mathbf{A}$ whose rows represent $f$ different biological functions, e.g., GO terms or KEGG pathways, and whose cells $a_{i,j}$ hold indicator variables whose value depends on whether an annotation exists between the function $i$ and genomic variable $j$. 
	\begin{equation} \label{eqn:A}
	 \mathbf{A} = \begin{bmatrix}
	  a_{1,1} & \cdots & a_{1,p} \\
	  \vdots  & \ddots & \vdots  \\
	  a_{f,1} & \cdots & a_{f,p} \\
	 \end{bmatrix}
	 , a_{i,j} = \indicator{\textit{variable $j$ has function $i$}}
	\end{equation}
	
\item \textit{Algorithm parameters:} 
The PCGSE method requires the specification of parameters that determine the gene-level statistic used to quantify association between genomic variables and PCs, whether transformations are applied to the gene-level statistics, the type of statistic calculated for each gene set and the method used to assess the significance of the gene set statistic given a competitive null hypothesis.

\end{enumerate}

\subsection{PCGSE algorithm}

Enrichment of the gene sets defined by $\mathbf{A}$ relative to one of the PCs of $\mathbf{X}$ is performed using the following sequence of steps:
\begin{enumerate}
\item Perform PCA on a standardized version of $\mathbf{X}$.
\item Compute gene-level statistics, $z_j, j=1,...,p$, for all $p$ genomic variables that quantify the association between the genomic variable and the PC.
\item (Optional) Transform the gene-level statistics.
\item Compute gene set statistics, $S_k, k=1,...,f$, for all $f$ gene sets defined by $\mathbf{A}$ using the gene-level statistics, $z_j$.
\item Determine the statistical significance of the gene set statistics according to a competitive null hypothesis.
\end{enumerate}
Each of these steps is explained in more detail in the sections \ref{sec:pca} thru \ref{sec:stat_sig} below. Note that steps 2 thru 5 have close parallels to modules in Ackermann and Strimmer's general modular framework for gene set enrichment analysis \citep{ackermann_general_2009}. 

\subsection{PCA for PCGSE}\label{sec:pca}

Because PCs are not invariant under scaling of the data \citep{jolliffe2002principal}, PCA is performed on a mean centered and standardized version of $\mathbf{X}$: 
\begin{equation} \label{eqn:std_X}
	\mathbf{\tilde{X}} = \begin{bmatrix}
	  \tilde{x}_{1,1} & \cdots & \tilde{x}_{1,p} \\
	  \vdots  & \ddots & \vdots  \\
	  \tilde{x}_{n,1} & \cdots & \tilde{x}_{n,p} \\
	 \end{bmatrix}
	 , \hat{x}_{i,j} = \frac{x_{i,j} - \bar{x}_{j}} {s_{x_{j}}}
\end{equation}
\noindent where $\bar{x}_{j}$ is the mean value of the $jth$ genomic variable computed over the $n$ samples and  $s_{x_{j}}$ is the sample standard deviation of the $jth$ genomic variable. The PC loading vectors and variances of $\mathbf{\tilde{X}}$ are thus the eigenvectors and eigenvalues of the sample correlation matrix, $\mathbf{S} = 1/(n-1) \mathbf{\tilde{X}}^T \mathbf{\tilde{X}}$, rather than the sample covariance matrix. 
For computational efficiency, the PCA solution is realized via the singular value decomposition (SVD) of $\mathbf{\tilde{X}}$:
\begin{equation} \label{eqn:SVD}
	\mathbf{\tilde{X}} = \mathbf{U} \boldsymbol{\Sigma} \mathbf{V}^T
\end{equation}
\noindent where the columns of $\mathbf{V}$ represent the PC loading vectors, the entries in the diagonal matrix $\boldsymbol{\Sigma}$ are proportional to the square roots of the PC variances and the columns of $\mathbf{U} \boldsymbol{\Sigma}$ are the PCs.

\subsection{Gene-level statistics} 

The PCGSE method supports the following gene-level statistics for quantifying the association between genomic variable $j$ and the target PC. These statistics are represented using the notation $z_j, j=1,...,p$.
\begin{itemize}
\item \textit{PC loading.} For genomic variable $j$ and target PC $m$, the gene-level statistic is element $v_{j,m}$ of matrix $\mathbf{V}$ from the SVD of $\mathbf{\tilde{X}}$ as defined in \eqref{eqn:SVD}.
\item \textit{Pearson correlation coefficient.} Where the correlation is computed between each genomic variable and the target PC. 
\item \textit{Fisher-transformed Pearson correlation coefficient.} This creates a statistic whose distribution is approximately $\mathcal{N}(0,1)$.
\end{itemize}
Because the Pearson correlation coefficients between genomic variables and PCs of the sample correlation matrix are proportional to the PC loadings (see \eqref{eqn:PC_cor} below), all of these gene-level statistics provide a measure of the correlation between genomic variables and PCs. 
\begin{align} \label{eqn:PC_cor}
	\begin{split}
	cor(\mathbf{\tilde{X}}, \mathbf{U} \boldsymbol{\Sigma}) &= cov(\mathbf{\tilde{X}}, \mathbf{U} \boldsymbol{\Sigma} \sqrt{1-n} \boldsymbol{\Sigma}^{-1}) \\
	 &= \frac{1}{n-1} (\mathbf{U} \boldsymbol{\Sigma} \mathbf{V}^T)^T \mathbf{U} \sqrt{1-n} \\
	 &= \frac{\sqrt{n-1}}{n-1} \mathbf{V} \boldsymbol{\Sigma} \mathbf{U}^T \mathbf{U} = \frac{1}{\sqrt{n-1}} \mathbf{V} \boldsymbol{\Sigma} 
	 \end{split}
\end{align}
\noindent where $\mathbf{U}, \boldsymbol{\Sigma}$ and $\mathbf{V}$ are from the SVD of $\mathbf{\tilde{X}}$ as specified in \eqref{eqn:SVD}.

The choice between the different gene-level statistics will be guided by the gene set statistic and significance testing method employed for PCGSE as well as computational constraints. For example, the added computational expense to generate z-statistics from correlation coefficients is motivated by parametric tests of the mean difference statistic, whereas, for rank sum tests, the PC loadings are sufficient.

\subsection{Transformation of gene-level statistics}

An absolute value transformation can optionally be applied to the gene-level statistics, i.e., $\tilde{z}_j = |z_j|$. Such a transformation gives the PCGSE method increased power to detect scale alternatives, i.e. gene sets that contain both significantly enriched and significantly repressed genomic variables, whereas the use of untransformed gene-level statistics provides better power against shift in location alternatives, i.e., gene sets containing genomic variables with a common direction of association \citep{Efron:2007uq}.

\subsection{Gene set statistics}

The PCGSE method supports two competitive gene set statistics for quantifying the association between gene set $k$ and a target PC. These statistics are represented using the notation $S_k, k=1,...,f$.

\subsubsection{Mean difference statistic} 

This statistic is computed as the standardized difference between the mean of the $z_j$ for genomic variables in the gene set and genomic variables not in the set and corresponds to $U_D$ in the notation of \cite{Barry2008}. Benefits of the mean difference statistic include its parametric null distribution and excellent power, relative to other gene set test statistics, for shift in location alternatives when using untransformed $z_j$ \citep{Efron:2007uq}. 
For gene set $k$, this statistic is defined as:
\begin{equation} \label{eqn:mean_diff_stat}
	S_k = \frac{\bar{z}_k - \bar{z}_{k^c}}{\sigma_p \sqrt{\frac{1}{m_k} - \frac{1}{p-m_k}}}
\end{equation}
\[
	m_k = \sum_{j=1}^{p} a_{k,j}, \bar{z}_k = \frac{\sum_{j=1}^{p} a_{k,j} z_j}{m_k}, \bar{z}_{k^c} = \frac{\sum_{j=1}^{p} !a_{k,j} z_j}{p-m_k}
\]
\noindent where $m_k$ is the number of genes in set $k$, $\bar{z}_k$ is the mean of the $z_j$ for members of gene set $k$, 
\noindent $\bar{z}_{k^c}$ is the mean of the $z_j$ for genes not in set $k$ and $\sigma_p$ is the pooled standard deviation of the $z_j$.

\subsubsection{Rank sum statistic} 

This statistic is computed as the standardized Wilcoxon rank sum statistic given the ranks of the $z_j$ for genomic variables in the set and genomic variables not in the set and corresponds to $U_W$ in the notation of \cite{Barry2008}. Benefits of the rank sum statistic include lack of distributional assumptions and robustness to outliers.  For gene set $k$, the Wilcoxon rank sum statistic is defined as the sum of the ranks of the gene-level statistic for all genomic variables belonging to gene set $k$ minus the minimum possible value for this sum of ranks:
\begin{equation} \label{eqn:rank_sum}
	W_k = \sum_{j =1}^{p} a_{k,j} \text{Rank}(z_{j}) - \frac{m_k (m_k +1)}{2}
\end{equation}
\noindent where $m_k = \sum_{j =1}^{p} a_{k,j}$, the size of gene set $k$. A version of this statistic that has an asymptotic $\mathcal{N}(0,1)$ distribution under the null can be generated as:
\begin{equation} \label{eqn:rank_sum_stat}
	S_k = \frac{W_k - \mu_{W_k}}{\sigma^2_{W_k}}
\end{equation}
\noindent where $\mu_{W_k} = (m_k(p-m_k))/2$ and $\sigma^2_{W_k} = (m_k(p-m_k)(m_k+1))/12$. 

Although it is possible to use any of the supported gene-level statistics with either of the gene set statistics, as discussed above, the mean difference statistic should be used with Fisher-transformed Pearson correlation coefficients and the rank sum statistic can be effectively used with just the PC loading elements.

\subsection{Gene set statistical significance}\label{sec:stat_sig}

To compute the statistical significance of the association between gene set $k$ and a target PC, the distribution of the gene set statistic $S_k$ must be calculated under the appropriate null hypothesis. The PCGSE approach supports three different methods (parametric, correlation-adjusted parametric and permutation) for computing the competitive null distributions of the mean difference and rank sum gene set statistics defined in \eqref{eqn:mean_diff_stat} and \eqref{eqn:rank_sum_stat}.

\subsubsection{Parametric tests} 

Under the competitive $H_0$ that the $z_j$ are independent and identically distributed, it is possible to determine the statistical significance of the association between each gene set and the target PC using a two-sided t-test for the mean difference statistic or a two-sided z-test for the rank sum statistic.  Both of these parametric tests fall into the class 1 test category as outlined in \cite{Barry2008} and are similar to the $Q_1$ test defined by \cite{Tian:2005zr}.

Under this $H_0$, the mean difference statistic defined in \eqref{eqn:mean_diff_stat} has a t-distribution with $p-2$ df and a two-sided t-test can therefore be used to determine statistical significance. For the rank sum statistic defined in \eqref{eqn:rank_sum_stat}, the asymptotic standard normal distribution under this $H_0$ can be used as the basis for a two-sided z-test. 

While it is often safe to assume a normal distribution for the $z_j$, especially after transformation, the $z_j$ will not be independent. Indeed, because the $z_j$ used with PCGSE are proportional to the PC loadings, they have an asymptotic multivariate normal distribution \citep{ANDERSON:1963fk}, assuming multivariate normality for the underlying genomic data, with significant correlation present between the loadings associated with the genes that have high pair-wise correlations \citep{jolliffe2002principal}. Because both the t-test for the mean difference statistic and the z-test for the rank sum statistic ignore this correlation between gene-level statistics, they will generate inflated type I error rates. These tests are therefore only supported by the PCGSE method for the purpose of comparative evaluation.

\subsubsection{Correlation-adjusted parametric tests} 

A computationally efficient approach for addressing correlation among the $z_j$ involves the use of correlation-adjusted parametric tests. Correlation-adjusted versions of the t-statistic associated with the mean difference statistic defined in \eqref{eqn:mean_diff_stat} and of the z-statistic associated with the rank sum statistic defined in \eqref{eqn:rank_sum_stat} were first discussed in the context of gene set testing by \cite{Barry2008}. Simplified versions of these correlation-adjusted statistics were later developed into the CAMERA method by \cite{Wu:2012fk}. Specifically, the approach taken by CAMERA assumes that correlation among the $z_j$ can be approximated by the correlation among the genomic variables (this is supported by results in \cite{Barry2008}), ignores all inter-gene correlation except the correlation among the members of the tested gene set and estimates a single average pair-wise correlation for gene set members using residuals from a linear regression.

The PCGSE method makes similar simplifying assumptions as those made by CAMERA , i.e., correlation between the $z_j$ can be approximated by correlation among the genomic variables, only gene set members have non-zero inter-gene correlation and all pair-wise correlations between gene set members are the same. An important difference between PCGSE and CAMERA is that PCGSE estimates the average inter-gene correlation directly from the sample correlation matrix. The correlation-adjusted mean difference statistic used by PCGSE is:

\begin{equation} \label{eqn:cor_adj_mean_diff_stat}
	S^{adj}_k = \frac{\bar{z}_k - \bar{z}_{k^c}}{\sigma_p \sqrt{\frac{\text{VIF}}{m_k} - \frac{1}{p-m_k}}}
\end{equation}
\noindent where $\text{VIF}$ (variance inflation factor)$ = 1 + (m_k - 1) \bar{\rho}_k$ and $\bar{\rho}_k$ is the average unbiased sample correlation between members of gene set $k$. Following \cite{Wu:2012fk}, this correlation-adjusted statistic has a t-distribution with $n-2$ df under $H_0$. Likewise the correlation-adjusted rank sum statistic is computed as:

\begin{equation} \label{eqn:cor_adj_rank_sum_stat}
	S^{adj}_k = \frac{W_k - \mu_{W_k}}{\sigma^2_{\text{VIF},W_k}}
\end{equation}
\noindent where $\sigma^2_{\text{VIF},W_k} = (m_k(p-m_k))/(2\pi) (sin^{-1} (1) + (p-m_k - 1) sin^{-1} (.5) + (m_k-1)(p-m_k-1)sin^{-1}(\bar{\rho}_k/2) + (m_k-1)sin^{-1}((\bar{\rho}_k+1)/2))$, as derived in \cite{Wu:2012fk} based on the formula in \cite{Barry2008}.

\subsubsection{Permutation test} 

The most common approach in the gene set testing literature for addressing correlation between gene-level statistics has been sample permutation. This approach, which corresponds to the class 2 test in \cite{Barry2008}, generates the null distribution of the gene set statistic via permutation of the outcome variable. For each permutation of the outcome variable, all gene-level statistics are recomputed to generate permutation statistics $z^*_j$ and then permutation gene set statistics $S^*_k$ are calculated using the $z^*_j$. The statistical significance for a given gene set $k$ is based on the proportion of all permutation $S^*_k$ more extreme than the observed $S_k$.  In standard gene set testing, permutation is applied to a clinical outcome variable, e.g., a case/control label. For PCGSE, permutation is applied to the elements of the target PC, i.e., the elements of one of the columns of $\mathbf{U} \boldsymbol{\Sigma}$. Because permutation is applied to the PC elements, this test can only be used with Pearson correlation coefficients or Fisher-transformed Pearson correlation coefficients as gene-level statistics.

A key assumption of the permutation null distribution is that the permuted values are i.i.d. Assuming the original $n$ observations of the p-dimensional random vector $\mathbf{x}$ are i.i.d, the elements of each PC will also be i.i.d., since each PC is a linear function of the original $\mathbf{x}$. Permutation of the PC elements therefore generates a valid permutation distribution for the mean difference and rank sum gene set statistics.

Because permutation tests handle correlation among the $z_j$ without attempting to estimate this correlation or make simplifying assumptions about the correlation structure, they are likely the most accurate of the statistical tests supported by PCGSE and are therefore used to evaluate the performance of the parametric and correlation-adjusted parametric tests. The exact permutation test was also used as a "gold-standard" in \cite{Zhou:2013ys}. Although they provide superior handling of inter-gene correlation, permutation tests do suffer from two important disadvantages relative to parametric tests: computational complexity and lower power to detect gene sets whose members all have a small common association with the outcome. Because of these disadvantages, correlation-adjusted parametric tests are preferred for most PCGSE applications. 

Another alternative to sample permutation testing that addresses the key challenge of computational complexity is the parametric approximation of the sample permutation distribution of gene-level score statistics developed by \cite{Zhou:2013ys}. Although the Zhou et al.'s beta distribution-based parametric approximations may be a useful option for the PCGSE method, it is not currently supported due to the lack of a parametric approximation for a directional, competitive gene set test statistic that is equivalent to the standardized mean difference statistic using untransformed directional gene-level test statistics. In \cite{Zhou:2013ys}, parametric approximations are only detailed for two self-contained gene set test statistics (sum of the score statistics and sum of the squares of the score statistics) and one non-directional competitive test statistic (a weighted sum of the squares of local score statistics).

\subsection{PCGSE output}

For each of the $f$ gene sets defined in $\mathbf{A}$ and each tested PC of $\mathbf{\tilde{X}}$, the PCGSE method outputs the observed value of the gene set test statistic, $S_k$, and a p-value representing the probability of encountering a gene set statistic as or more extreme than then the observed $S_k$ under the appropriate competitive null hypothesis.
	
\subsection{PCGSE evaluation}

\subsubsection{Evaluation using simulated gene sets and simulated data.}\label{sec:simulation_methods}

As a simple example, the PCGSE method was used to compute the statistical association between 20 disjoint gene sets, each of size 10, against the PCs of 100 simulated gene expression datasets each comprised by 50 independent observations of a 200-dimensional random vector simulated according to a multivariate normal distribution $\sim \text{MVN}(\boldsymbol{\mu}, \boldsymbol{\Sigma})$. The population covariance matrix, $\boldsymbol{\Sigma}$, was generated as: $\boldsymbol{\Sigma} = \lambda_1 \boldsymbol{\alpha}_1 \boldsymbol{\alpha}_1^T + \lambda_2 \boldsymbol{\alpha}_2 \boldsymbol{\alpha}_2^T + \lambda_d \boldsymbol{I}$, where $\lambda_1=$ 2, $\lambda_2=$ 1,$\lambda_d=$ 0.1, $\boldsymbol{\alpha}_1$ is a 200-dimensional vector with all elements equal to $0$ except for the first 10 which were set to $\sqrt{.1}$, $\boldsymbol{\alpha}_2$ is a 200-dimensional vector with all elements equal to $0$ except for the second 10 which were set to $\sqrt{.1}$. Figure \ref{fig:sim_model} shows the variance and loadings for both population and sample PCs simulated according to this model.

For this simulated example, the PCGSE method was executed using the Fisher-transformed Pearson correlation coefficient between each variable and each PC as the gene-level test statistic with the standardized mean difference, as defined in \eqref{eqn:mean_diff_stat}, as the gene set test statistic. The statistical significance of the association between each of the 20 simulated gene sets and each PC was computed using all supported tests described in Section \ref{sec:stat_sig}: parametric, correlation-adjusted parametric and permutation. For the standardized mean difference gene set statistic, these tests were realized by a two-sided t-test, a correlation-adjusted two-sided t-test and a two-sided permutation test based on permutation of the PC elements, respective. 

Because the true association was known between simulated gene sets and the PCs of the simulated data, it was possible to compute contingency table statistics. In this case, the type I error rates for the different statistical testing methods were computed for gene set 2 relative to PC 1 and for gene set 1 relative to PC 2, both cases with no true association. 

\subsubsection{Evaluation using Spellman et al. $\alpha$ factor-synchronized yeast gene expression data and yeast cell cycle gene sets.}\label{sec:cellcycle_methods}

The PCGSE method was used to compute the statistical association of the yeast cell cycle gene sets defined by Spellman \citep{spellman_comprehensive_1998} relative to the first three PCs of a specially processed version of the $\alpha$ factor-synchronized yeast gene expression data collected by \cite{spellman_comprehensive_1998} and re-examined by \cite{alter_singular_2000}. 

Both the $\alpha$ factor-synchronized data and yeast cell cycle gene sets were downloaded from the supplementary material website for \cite{alter_singular_2000}. To support comparison against the results reported in \cite{alter_singular_2000}, PCA was performed on a version of the gene expression data that was specially processed according to the steps outlined in \cite{alter_singular_2000} so that the first three PCs were identical to the first three so-called eigengenes. Figure \ref{fig:cellcycle} is a reproduction of Figure 5 from \cite{alter_singular_2000} with the value of the first three PCs of the specially processed gene expression data (i.e., the eigengenes) shown relative to the 22 $\alpha$ factor arrays.

The PCGSE method was executed on the Spellman et al. data and gene sets using the Fisher-transformed Pearson correlation coefficient between each gene and each PC as the gene-level test statistic and the standardized mean difference, as defined in \eqref{eqn:mean_diff_stat}, as the gene set statistic. Similar to the simulation example outlined in Section \ref{sec:simulation_methods}, the statistical significance of the gene set statistic was computed using all supported tests described in Section \ref{sec:stat_sig}. 

\subsubsection{Evaluation using MSigDB C2 v4.0 gene sets and Armstrong et al. leukemia gene expression data.}\label{sec:leukemia_methods}

The PCGSE method was also used to compute the statistical association between the MSigDB C2 v4.0 gene sets and the first 3 PCs of the leukemia gene expression data \citep{Armstrong:2002fk} used in the 2005 GSEA paper \citep{subramanian_gene_2005}.  

The MSigDB C2 v4.0 cancer modules and collapsed leukemia gene expression data were both downloaded from the MSigDB repository. With a minimum gene set size of 15 and maximum gene set size of 200, 3,076 gene sets out of the original 4,722 were used in the analysis. Similar to the simulation example outlined in Section \ref{sec:simulation_methods} and the yeast cell cycle example outlined in Section \ref{sec:cellcycle_methods}, the PCGSE method was executed using the Fisher-transformed Pearson correlation coefficient between each genomic variable and each PC as the gene-level test statistic and the standardized mean difference, as defined in \eqref{eqn:mean_diff_stat}, as the gene set test statistic. The statistical significance of the association between each of the MSigDB C2 gene sets and each of the first 3 PCs of the standardized leukemia gene expression data was computed using all supported tests described in Section \ref{sec:stat_sig}.

The enrichment of the MSigDB C2 gene sets was also computed relative to the acute myeloid leukemia (AML) versus acute lymphoblastic leukemia (ALL) phenotype using the GSA method \citep{Efron:2007uq} with the restandardized mean statistic and 10,000 permutations. For each of the first three PCs and each of the PCGSE methods for computing statistical significance of the standardized mean difference gene set statistic, the Spearman correlation coefficient was computed between PC gene set enrichment p-values and phenotype enrichment p-values. For PC 2, for which the PC and phenotype gene set enrichment p-values were highly correlated, contingency table statistics were computed measuring how well PCGSE was able to identify MSigDB C2 gene sets significantly associated with the AML/ALL phenotype.


\section{Results and Discussion}

\subsection{Enrichment of simulated gene sets relative to PCs of simulated data}\label{sec:sim_results}

\begin{figure}[!ht]
\centering
\includegraphics[width=.75\textwidth]{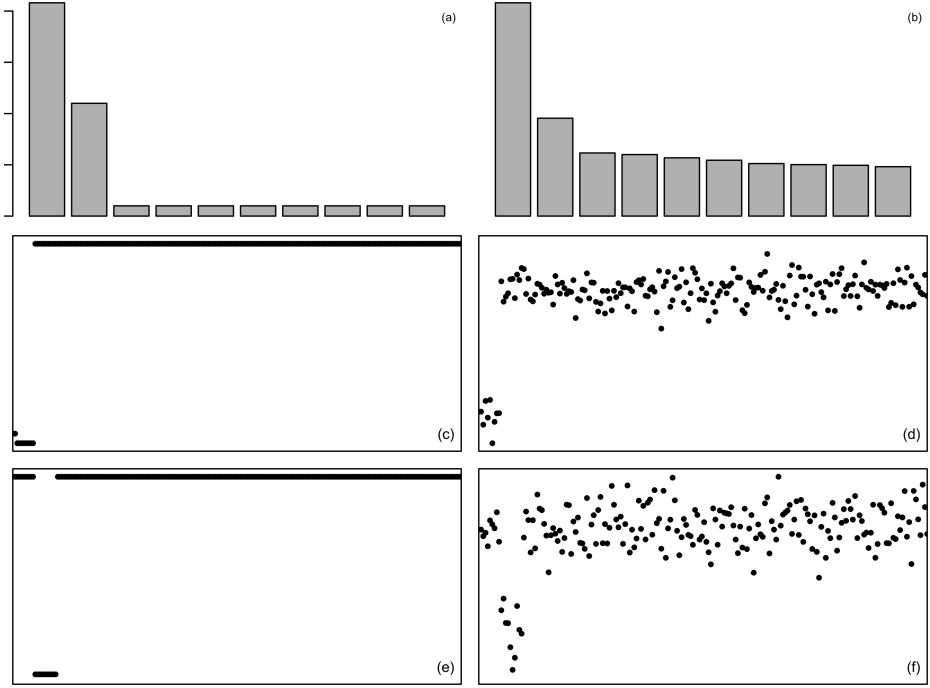}
\caption{
Simulation model.
Variances and loadings for the principal components a 200-dimensional population covariance matrix, $\boldsymbol{\Sigma}$, and the sample covariance matrix estimated from n=50 independent observations of the random vector $\mathbf{x} \sim MVN(\mathbf{0}, \boldsymbol{\Sigma})$ where $\boldsymbol{\Sigma}$ is generated according to the model outlined in Section \ref{sec:simulation_methods}.
Variances for the first ten population PCs are shown in plot \textbf{(a)} and loadings for the first two population PCs are shown in plots \textbf{(c)} and \textbf{(e)}. Plots \textbf{(b)}, \textbf{(d)} and \textbf{(f)} show the corresponding variances and loadings for the sample PCs of a single simulated dataset.
}
\label{fig:sim_model}
\end{figure}

\begin{figure}[!ht]
\centering
\includegraphics[width=.75\textwidth]{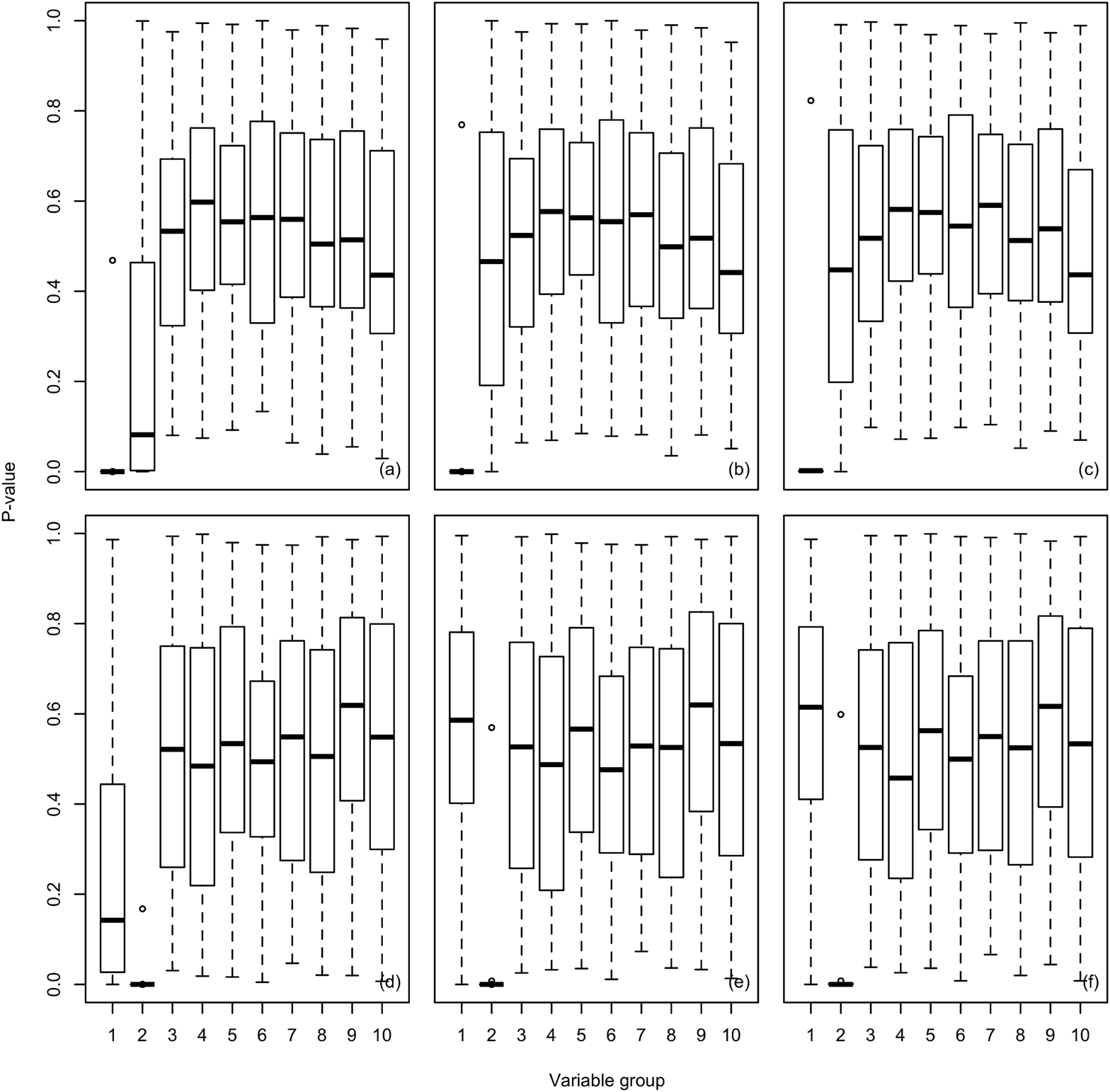}
\caption{
Simulation results.
Distribution of PCGSE-computed enrichment p-values for the first 10 of 20 simulated gene sets relative to the first 2 PCs of 100 datasets simulated according to the model described in Section \ref{sec:simulation_methods} and illustrated in Figure \ref{fig:sim_model}. The PCGSE method was executed using parameter settings outlined in Section \ref{sec:simulation_methods}.
For all displayed results, PCGSE was executed using the Fisher-transformed Pearson correlation coefficient between each genomic variable and each PC as the gene-level test statistic and the standardized mean difference as the gene set test statistic. 
Plots \textbf{(a)}, \textbf{(b)} and \textbf{(c)} display the distribution of enrichment p-values for the first 10 gene sets relative to the first PC of all simulated data sets.
In plots \textbf{(d)}, \textbf{(e)} and \textbf{(f)}, enrichment p-values computed relative to the second PC are displayed.
For plots \textbf{(a)} and \textbf{(d)}, the p-values were computed using a two-sided t-test on the standardized mean difference gene set test statistic,
for plots \textbf{(b)} and \textbf{(e)}, the p-values were computed using a two-sided correlation-adjusted t-test and, 
for plots \textbf{(c)} and \textbf{(f)}, the p-values were computed using a two-sided permutation test.
}
\label{fig:sim_results}
\end{figure}

According to the population covariance matrix, $\boldsymbol{\Sigma}$, used to simulate the 100 datasets, only the first gene set should be significantly enriched on the first PC and only the second gene set should be significantly enriched on the second PC. This relationship can seen easily in the loading values for population PCs 1 and 2 as shown in Figure \ref{fig:sim_model} plots \textbf{(c)} and \textbf{(e)}. The significant loading of gene set 2 on PC 2, however, will result in a high pair-wise correlation between the PC loadings for gene set 2 members on PC 1. The fact that high loadings on one PC result in correlation among the PC loadings on other PCs follows from the formula for the asymptotic distribution of the PC loadings for MVN data \citep{ANDERSON:1963fk}:

\begin{equation}\label{eq:v_asym}
	\mathbf{v}_j \sim \mathcal{N}(\boldsymbol{\alpha}_j, \mathbf{T}_j), j=1,...,p
\end{equation}
\begin{equation}\label{eq:T}
	\mathbf{T}_j = \frac{\lambda_j}{n-1} \sum_{k=1, k \neq j}^{p} \frac{\lambda_k \boldsymbol{\alpha_k \alpha_k}^T}{(\lambda_k-\lambda_j)}
\end{equation}
where 
\begin{itemize}
\item $p$ is fixed and $n \to \infty$.
\item $\lambda_j$ an eigenvalue of the population covariance matrix, $\lambda_1 > \lambda_2 > ... > \lambda_p$.
\item $\boldsymbol{\alpha}_j$ is an eigenvector of the population covariance matrix.
\end{itemize}

The gene-level test statistics computed for gene set 2 on PC 1 and for gene set 1 on PC 2 will therefore have a non-zero average pair-wise correlation. The impact of this correlation between the gene-level test statistics can be seen in the PCGSE results shown in Figure \ref{fig:sim_results}. The unadjusted t-test uses an incorrectly small variance for the mean difference statistic and, as expected, generates the high type I error rate of $0.42$ given a nominal $\alpha$ of $0.05$ for gene set 2 relative to PC 1 and $0.3$ for gene set 1 relative to PC 2. The correlation-adjusted two-sided t-test and the two-sided permutation test are much more successful at controling the type I error rate. For PC 1 and gene set 2, the type I error rate was $0.13$ for both correlation-adjusted t-test and the permutation test. For PC 2 and gene set 1, the type I error rate was
$0.06$ for both the correlation-adjusted t-test and the permutation test. For this example, all gene set testing methods were able to correctly reject the null hypothesis for almost all cases where the gene set had a true association with the PC, e.g., gene set 1 relative to PC 1 and gene set 2 relative to PC 2.

Although based on a simple two-factor MVN model, this simulation example demonstrates the importance of controlling for correlation between gene-level test statistics when computing PC gene set enrichment. Tests which assume independence among the statistics that quantify the association between genes and PCs, such as a two-sample t-test, Fisher's exact test or a gene permutation test, will underestimate the variance of the gene set test statistic and therefore reject too many null hypotheses. This example also shows that the correlation-adjusted t-test can achieve enrichment sensitivity and specificity comparable to a sample permutation test with a significantly lower computational burden.

PCGSE was was computed for this simulation example using the standardized rank sum as the gene set statistic. 
The results for the standardized rank sum statistic using parametric and permutation tests are similar to those for the standardized mean difference statistic. Although the correlation-adjusted z-test based on the standardized rank sum statistic has an improved type I error rate, it has an inflated type II error rate, i.e., it rejects too few null hypotheses when the association is true. The inflated type II error rate for the correlation-adjusted rank sum z-test is likely due to an overestimated VIF under the alternative hypothesis, as computed by the equation from \cite{Barry2008}. Therefore, in cases where a rank sum gene set statistic is motivated, PCGSE should be performed using a permutation test and not using the more efficient correlation-adjusted z-test.

\subsection{Enrichment of Spellman et al. yeast cell cycle gene sets relative to the PCs of Spellman et al. $\alpha$-synchronized yeast gene expression data}\label{sec:cellcycle_results}

\begin{figure}[!ht]
\centering
\includegraphics[width=.75\textwidth]{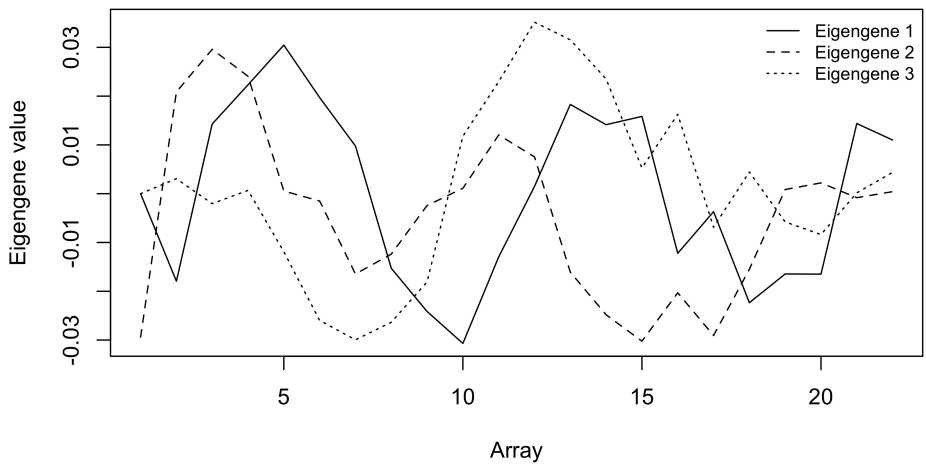}
\caption{
Reproduction of Figure 5 from \cite{alter_singular_2000} that displays the value of first three PCs of the specially processed gene expression matrix, i.e., the "eigengenes", over the 22 $\alpha$ factor-synchronized arrays. According to \cite{alter_singular_2000}, the following approximate mapping held between arrays and cell cycle phases: 1 (NA), 2 (M/G1), 3,4 (G1), 5,6 (S), 7,8 (S/G2), 9 (G2/M), 10,11 (M/G1), 12,13 (G1), 14,15 (S), 16 (S/G2), 17,18 (G2/M), 19,20 (S/G1), 21,22 (G1)
}
\label{fig:cellcycle}
\end{figure}

\begin{table}[!t]\label{table:cellcycle}
\centering
\begin{tabular}{ l c c c c c c } 
\hline
& \multicolumn{2}{c}{T-test} & \multicolumn{2}{c}{Cor-adj t-test} & \multicolumn{2}{c}{Perm} \\
& PC 1 & PC 2 & PC 1 & PC 2 & PC 1 & PC 2 \\ 
\hline
$M/G_1$ & 0.68 & \textbf{1.2e-12} & 0.94 & 0.18 & 0.94 & 0.22 \\ 
$G_1$ & \textbf{3.5e-130} & \textbf{1e-35} & \textbf{0.023} & 0.23 & \textbf{0.024} & 0.35 \\ 
$S$ & \textbf{1e-10} & \textbf{0.0074} & 0.2 & 0.59 & 0.2 & 0.62 \\ 
$S/G_2$ & 0.27 & \textbf{4.6e-06} & 0.86 & 0.45 & 0.87 & 0.47 \\ 
$G_2/M$ & \textbf{8.3e-38} & 0.068 & 0.07 & 0.79 & \textbf{0.048} & 0.81 \\
\hline
\end{tabular}
\caption{PCGSE computed enrichment p-values for the \cite{spellman_comprehensive_1998} yeast cell cycle gene sets relative to the first two PCs of the \cite{spellman_comprehensive_1998} $\alpha$ factor-synchronized gene expression data processed using the steps outlined in \cite{alter_singular_2000}. PCGSE was executed using Fisher transformed Pearson correlation coefficients between genes and PCs as gene-level test statistics. Significance of the standardized mean difference gene set statistic was computed using either a two-sided t-test, a correlation-adjusted two-sided t-test or a two-sided permutation test. Unadjusted p-values less than 0.05 are displayed in bold.}
\end{table}

The \cite{spellman_comprehensive_1998} $\alpha$ factor-synchronized gene expression data was selected for PCGSE analysis because it is well known, has been widely reanalyzed, is easily accessible and has a spectra with a published biological interpretation. In particular, the reanalysis by \cite{alter_singular_2000} was one of the first to illustrate that the spectra of gene expression data can represent important biological features, in this case phases of the yeast cell cycle. In \cite{alter_singular_2000}, the authors provided a qualitative interpretation of the first two eigengenes in terms of the yeast cell cycle by examining the correlation between the eigengenes and genes known to be active during different cell cycle phases, as defined by Spellman et al.'s yeast cell cycle gene sets. Alter et al. concluded that the first eigengene was correlated with genes that peak late in cell cycle phase $G_1$ and early in phase $S$ and was anticorrelated with genes that peak late in cell cycle phase $G_2/M$ and early in phase $M/G_1$. Alter et al. also concluded that the second eigengene was correlated with genes that peak late in cell cycle phase $M/G_1$ and early in phase $G_1$ and was anticorrelated with genes that peak late in phase $S$ and early in phase $S/G_2$.

Table \ref{table:cellcycle} contains p-values representing the statistical significance of the association between each of the Spellman et al. yeast cell cycle gene sets and the first two PCs of a specially processed version of the Spellman et al. gene expression data. As described in Section \ref{sec:cellcycle_methods}, this special processing ensured that the PCs were identical to the eigengenes analyzed in \cite{alter_singular_2000}. When a unadjusted two-sided t-test was used to determine the statistical significance of the standardized mean difference gene set statistic, the gene sets corresponding to cell cycles $G_1$, $S$ and $G_2/M$ were all highly significantly associated with PC 1 and the gene sets corresponding to M/$G_1$, $G_1$, $S$ and $S/G_2$ were all significantly associated with PC 2.  However, when either a correlation-adjusted two-sided t-test or two-sided permutation was used to determine the statistical significance of the standardized mean difference set statistic, PC 1 only had a significant association with the gene set corresponding to phase $G_1$ (with a marginally significant association with phase $G_2/M$) and none of the cell cycle gene sets were significantly associated with PC 2. 

Comparing the output from PCGSE with the analysis in \cite{alter_singular_2000}, the results from the unadjusted two-sided t-test align closely with the qualitative conclusions of Alter et al. The output from the correlation-adjusted t-test and permutation test, although generally in agreement for PC 1, are in direct contract with Alter et al. regarding PC 2, finding no cell cycle association.
The agreement between Alter et al. and the unadjusted t-test results is expected since the authors had based their analysis simply on a qualitative inspection of the gene-level correlations without a more formal test of a gene set test statistic that took account of the correlation between the gene-level test statistics associated with each cell cycle phase. The fact that the more accurate PCGSE methods failed to find an association between PC 2 and the cell cycle gene sets indicates that the originally published association in \cite{alter_singular_2000} was a false positive due to either the high inter-gene correlation present among the members of these sets or the selective examination by Alter et al. on a subset of the genes in each of the cell cycle gene set with a common direction of association with the eigengene. In the later case, it is likely that a gene set statistic such as the maxmean \citep{Efron:2007uq} would identify significant cell cycle enrichment for the second eigengene.

This example highlights the importance of using formal statistical methods for gene set testing when attempting to interpret the PCs of genomic data in terms of gene sets. Such gene set testing methods must specifically account for the correlation between gene-level test statistics. 

\subsection{Enrichment of MSigDB C2 v4.0 gene sets relative to PCs of Armstrong et al. leukemia gene expression data}\label{sec:leukemia_results}

\begin{figure*}[!ht]
\centering
\includegraphics[width=.9\textwidth]{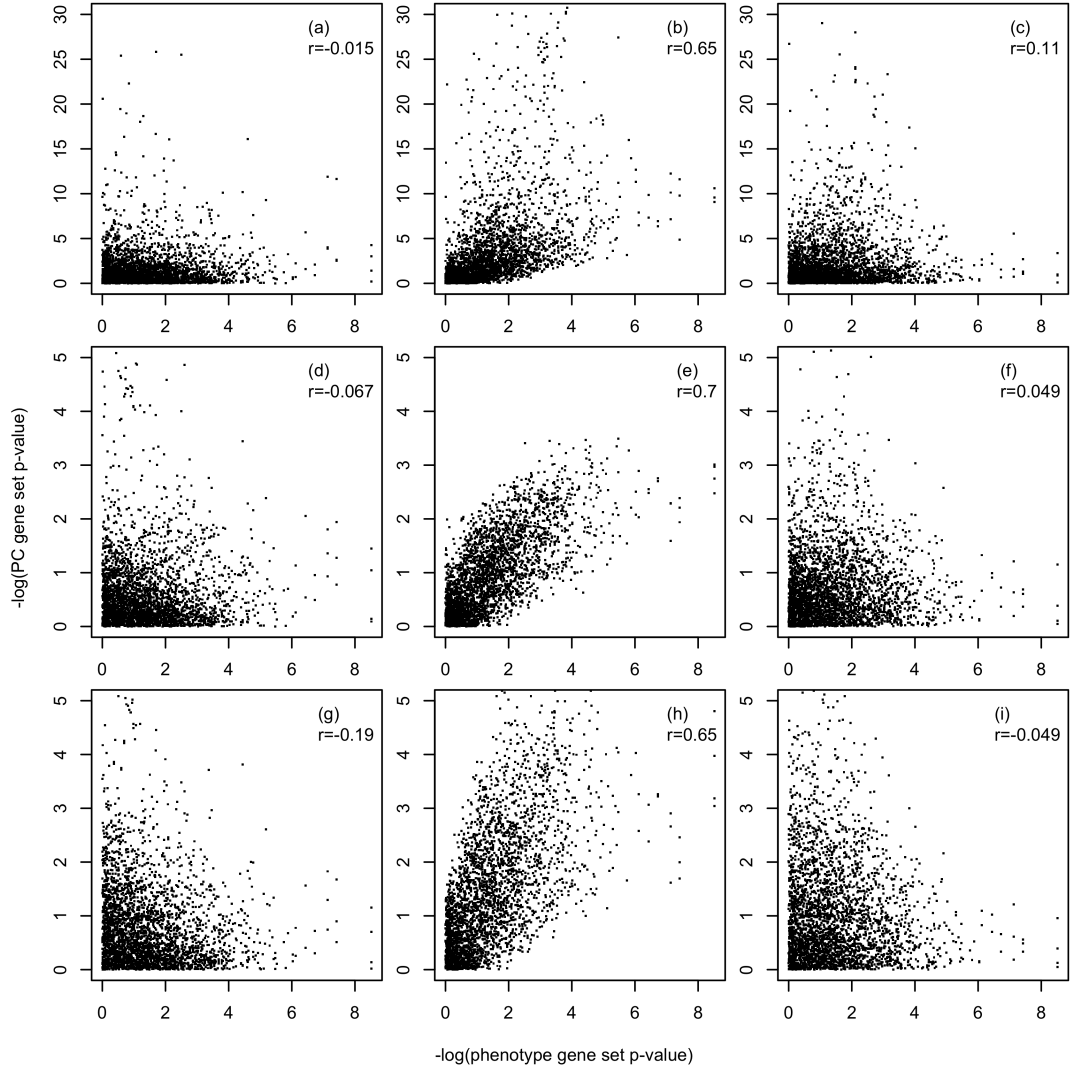}
\caption{
Scatter plots showing the association between phenotype gene set enrichment p-values and PC gene set enrichment p-values for the \cite{Armstrong:2002fk} leukemia gene expression data, AML/ALL phenotype, MSigDB C2 v4.0 gene sets and first three PCs. Both phenotype and PC gene set enrichment p-values were computed as outlined in Section \ref{sec:leukemia_methods}. The Spearman correlation coefficient between phenotype and PC gene set enrichment p-values is displayed in each plot. Plots \textbf{(a)-(c)} show the association between phenotype and PC gene set enrichment p-values for PCs 1 through 3 with the PC enrichment p-values computed using a two-sided t-test on the standardized mean difference gene set statistic. For plots \textbf{(d)-(f)}, the PC gene set enrichment p-values were computed using a correlation-adjusted two-sided t-test and, for plots \textbf{(g)-(i)}, the PC gene set enrichment p-values were computed using the permutation distribution of the gene set statistic.}
\label{fig:leukemia}
\end{figure*}
 
The classic \cite{Armstrong:2002fk} leukemia gene expression dataset is another excellent example of a case where the genomic patterns associated with an interesting phenotype have a clear representation in the spectral structure of the data. For the Armstrong et al. data, the second PC of the gene expression data is strongly associated with the AML versus ALL status of the subjects. Use of the Armstrong et al. gene expression data and MSigDB C2 v4.0 gene sets for evaluation of PCGSE was also motivated by the extensive use of this dataset and gene set collection in the gene set enrichment literature (e.g., \cite{subramanian_gene_2005}) and easy accessibility from the MSigDB repository, factors that will facilitate interpretation and replication of the reported PCGSE results by other researchers.
 
Figure \ref{fig:leukemia} shows the association between phenotype and PC gene set enrichment p-values for the MSigDB C2 v4.0 gene sets, the AML versus ALL phenotype and the first three PCs of the Armstrong et al. leukemia gene expression data. Each of the columns in the multi-plot corresponds to results for one PC and each row corresponds to one of the three different statistical tests supported by PCGSE on the standardized mean difference gene set statistic (i.e., t-test, correlation-adjusted t-test and permutation test). The association between PC 2 and the AML versus ALL phenotype can be clearly seen in Figure \ref{fig:leukemia} plots \textbf{(b)}, \textbf{(e)} and \textbf{(h)}. For all three PCGSE methods, the PC enrichment p-values for the MSigDB C2 v4.0 gene sets are highly correlated with the enrichment p-values computed for these gene sets relative to the AML versus ALL phenotype.

Similar to the PCGSE results outlined in previous sections on simulated data and yeast gene expression data, the unadjusted two-sided t-test on the standardized mean difference gene set statistic generates PC gene set enrichment p-values that are substantially lower than the enrichment p-values output by either the correlation-adjusted t-test or the permutation test. Although the true enrichment status of the MSigDB C2 v4.0 gene sets relative to the PCs of the \cite{Armstrong:2002fk} gene expression data is unknown, the phenotype enrichment results can be used as a proxy for the true gene set association with PC 2 under the assumption that this PC captures the AML versus ALL signal. If gene sets with a phenotype enrichment significance at or below 0.05 are considered AML/ALL markers, the PCGSE method is able to correctly identify these gene sets via enrichment relative to PC 2 with an area under the receiver operator characteristic curve (AUC) of 0.85 for the t-test results displayed in plot \textbf{(b)}, an AUC of 0.89 for the correlation-adjusted t-test results displayed in plot \textbf{(e)} and an AUC of 0.85 for the permutation test results displayed in plot \textbf{(h)}. Considering identification of AML/ALL-associated gene sets via PC enrichment using just $\alpha=$0.05, the PCGSE method has a positive predictive value of 0.26 for the t-test results displayed in plot \textbf{(b)}, 0.91 for the correlation-adjusted t-test results displayed in plot \textbf{(e)} and 0.46 for the permutation test results displayed in plot \textbf{(h)}. 

PCGSE analysis of the MSigDB C2 v4.0 gene sets and \cite{Armstrong:2002fk} leukemia gene expression data illustrates the biological motivation for PC gene set enrichment and demonstrates the superior performance of the computationally efficient correlation-adjusted t-test relative to either an unadjusted t-test or permutation test. 

\section{Conclusion}

Although principal component analysis is widely used for the dimensional reduction of biomedical data, with applications in visualization, clustering and regression, interpretation of PCA-based models remains challenging. While rotation methods and sparse PCA techniques can generate approximate PCs with few non-zero loadings that support interpretation in terms of individual variables, these approaches will perform poorly on genomic data in which important biological signals are defined by the collective action of groups of functionally related genes. Although gene set testing methods have been widely applied to analyze the association between gene sets and clinical phenotypes, such variable group testing methods have seen little application for testing the association between gene sets and the spectra of genomic data.To address the challenge of PC interpretation for genomic data and support the interpretation of genomic PCs in terms of functional gene sets, we have developed the principal component gene set enrichment (PCGSE) method. PCGSE performs a two-stage competitive gene set test using the correlation between each gene and each PC as the gene-level test statistic with flexible choice of both the gene set test statistic and the method used to compute the null distribution of the gene set statistic. To facilitate use of the PCGSE method by other researchers, an implementation of the technique is available as an R package from CRAN. On both simulated gene sets with simulated data and on curated gene sets with real gene expression data, a computationally efficient version of the PCGSE method based on a correlation-adjusted two-sided, two-sample t-test has been shown to accurately compute the statistical association between gene sets and the PCs of genomic data.

\section*{Acknowledgement}

\paragraph{Funding:} National Institutes of Health R01 grants LM010098, LM011360, EY022300, GM103506 and GM103534.
\paragraph{Conflict of Interest:} None declared.


\bibliographystyle{natbib}
\bibliography{PCGSE}

\end{document}